\documentclass[preprint,12pt]{elsarticle}

\usepackage{subfigure}
\usepackage{graphicx}% Include figure files
\usepackage{dcolumn}% Align table columns on decimal point
\usepackage{bm}% bold math
\usepackage{appendix}
\usepackage{xcolor}
\usepackage[dvipsnames]{xcolor}
\usepackage{amsmath}
\usepackage{amssymb}
\usepackage{hyperref}
\usepackage{cleveref}
\usepackage{tikz}
\usetikzlibrary{decorations.pathreplacing,arrows.meta}
\colorlet{streamA}{blue!16}\colorlet{streamB}{orange!22}
\colorlet{streamAd}{blue!55!black}\colorlet{streamBd}{orange!70!black}
\usepackage{lmodern}

\newcommand{\vpar}{v_\parallel}
\newcommand{\modB}{\|B\|}

\newcounter{bla}

 \journal{Computer Physics Communications}
\begin{document}

\begin{frontmatter}

\title{CATAPULT: A CUDA-Accelerated Timestepper for Alpha Particles Using Local Tricubics}

\author[inst1]{Michael Czekanski \corref{author}}
\author[inst2]{Alexey R. Knyazev}
\author[inst3]{David Bindel}
\author[inst2]{Elizabeth J. Paul}

\cortext[author] {Corresponding author.\\\textit{E-mail address:} mc2589@cornell.edu}

\affiliation[inst1]{organization={Department of Statistics and Data Science, Cornell University},
            city={Ithaca}, 
            state={NY},
            country={USA}}

\affiliation[inst2]{organization={Department of Applied Physics and Applied Mathematics, Columbia University},
            city={New York}, 
            state={NY},
            country={USA}}

\affiliation[inst3]{organization={Department of Computer Science, Cornell University},
            city={Ithaca}, 
            state={NY},
            country={USA}}

\begin{abstract}
We introduce a CUDA-Accelerated Timestepper for Alpha Particles Using Local Tricubics (CATAPULT) for use in Monte Carlo calculations of alpha particle confinement in stellarators.
Our GPU implementation is significantly faster than CPU-parallelized implementations of the same field representation such as SIMSOPT, and handles both equilibrium magnetic fields and Shear Alfven Waves.
We test our implementation on several example stellarators to exhibit both the speed and correctness of our code. The source code is included in the \textbf{firm3d} Python package.
\\

\noindent \textbf{CUDA-Accelerated Timestepper for Alpha Particles Using Local Tricubics}

\begin{small}
\noindent
{\em Program Title: CATAPULT}                                          \\
{\em CPC Library link to program files:} (to be added by Technical Editor) \\
{\em Developer's repository link: https://github.com/ColumbiaStellaratorTheory/firm3d/} \\
{\em Licensing provisions:} MIT  \\
{\em Programming language: Python, C++, CUDA}                                   \\
{\em Nature of problem:
Alpha particle tracing is a necessary component of viable stellarator design to understand particle confinement, wall loading from lost particles etc. Desirable stellarators, with low alpha particle losses, in particular require tracing many particles to better understand loss distributions. This requires a Monte Carlo scheme paired with an efficient ODE timestepper to compute particle trajectories.}\\
{\em Solution method:
We implement the Dormand-Prince 5 timestepper with a tricubic magnetic field representation for tracing alpha particles implemented in C++ and CUDA. This improves estimates of statistics of alpha particle transport by greatly improving particle throughput, reducing stochastic Monte Carlo error in our estimates of these statistics. }\\
  %Describe the method solution here.
{\em Additional comments including restrictions and unusual features : Our code can be used to trace particles in both Boozer and Cartesian coordinates. In Boozer coordinates we can trace particles in both equilibrium magnetic fields and in the presence of Shear Alfven Waves.}\\
  %Provide any additional comments here.
   \\
\end{small}
   \end{abstract}
\end{frontmatter}

\section{Introduction}

Magnetic confinement devices for fusion-relevant plasmas require the confinement
of high-temperature ions, keeping energy within the plasma and maintaining
temperatures necessary for subsequent fusion reactions.
Evaluation of devices requires understanding alpha particle 
confinement, as explored by Bonofiglo et al. \citep{bonofiglo2025fast} and directly
optimized by Bindel et al. \citep{bindel2023direct}.
In stellarators, this evaluation is generally performed with Monte Carlo methods paired with ODE
timesteppers of different varieties, as implemented in SIMSOPT \cite{landreman2021simsopt},
SIMPLE \cite{albert2020accelerated}, ORBIT \cite{white_hamiltonian_1984}, ASCOT \cite{hirvijoki2014ascot},
and TAPaS \cite{zarzoso2022transport}, and others.
Particles are sampled from the fusion birth distribution and traced until they leave the 
last closed flux surface or reach some maximum time $T$. 
As in other areas of plasma research, GPU acceleration can improve Monte Carlo estimation, as seen in particle-in-cell methods \cite{burau2010picongpu, xiong2024gpic}.
We review the motion of these particles and the Monte Carlo methods used for evaluation
before introducing a new implementation of the ODE timestepper: a CUDA-Accelerated
Timestepper for Alpha Particles Using Local Tricubics (CATAPULT). 
We present the runtime scaling of CATAPULT and describe future uses of this code, including the presence
of Shear Alfven Waves in the magnetic field.
Our code has higher throughput in higher fidelity simulations, reducing Monte Carlo error in evaluating stellarator designs.

We assume the deterministic particle dynamics are given by 
the evolution map $f(x,t)$ where $x$ is the initial point in phase space (position and parallel velocity with fixed velocity). $f(x,t)$ is the position in phase space at time t given a particle position of $x$ at time $0$.
The dynamics depend on the magnetic field, which may be time varying.
For a given magnetic field, we are interested in estimating
\begin{align} \label{eq:qoi}
\phi = \int_\Omega g(y(x)) \mu(x) dx
\end{align}
where $T$ is the maximum time particles are observed, $\tau$ is a stopping condition on the trajectory
(e.g. the time it leaves the last closed flux surface), $x$ is the initial position in phase space $\Omega$,
$\mu$ is a birth distribution over phase space, and $g$ is some measurement at the last point we observe a particle.
Then $t_{\max}(x) := \min(\tau(x), T)$ is the last point we need to observe a particle and $y(x) := f(x, t_{\max(x)})$.
In summary, $\mu$ describes the distribution of initial conditions, $y$ describes the dynamics, and $g$ is the quantity of interest.

The choice of $g$ depends on the problem of interest. 
For example, if we are interested in the proportion of particles that are lost, $g$ is the indicator $g(y) = \mathrm 1_{y \text{ outside plasma}}(y)$.
Similar functions $g$ can be chosen to calculate energy loss (see Bindel et al. \cite{bindel2023direct}),
wall-loads from alpha particles, and other quantities of interest.

The integral in \Cref{eq:qoi} is solved by sampling $n$ initial conditions $\{X_i : i \in [n]\}$ independently from $\mu$
and computing an approximation $Y_i := \hat y(X_i)$ of the last location of interest
for a particle born at $X_i$. 
Our Monte Carlo estimate is then
\begin{align}
\hat \phi = \frac{1}{n} \sum_{i=1}^n g(Y_i).
\end{align}

Our evaluation of $g$ is exact, and $\hat y$ is a numerical approximation to $y$ which yields two sources 
of error: numerical error in the timestepping and $O(n^{-1/2})$ stochastic error from our sampling scheme.
In this work, we focus on the latter.
Existing Monte Carlo calculations in SIMSOPT and SIMPLE target stellarator geometry but are limited in throughput by CPU-only implementations. ORBIT and ASCOT have GPU implementations \cite{clauser2019focus,peybernes2025gpu}, FOCUS has a GPU full orbit implementation \cite{clauser2019focus},  but our work is the first to use the SIMSOPT representation of magnetic fields in stellarators with guiding center tracing in Boozer and Cartesian coordinates with and without Shear Alfven Waves.
We obtain an order of magnitude improvement in throughput over SIMSOPT parallelized on 128 CPU threads, which corresponds to better estimates of loss statistics and resolution of alpha particle behavior.

\section{Particle orbit model}

Drift motion of a particle $i$ with mass $M_{\rm i}$ and charge $q_{\rm i}$, moving in an equilibrium magnetic field $\mathbf{B}_0=\nabla\times\mathbf{A}_0$ with vector potential $\mathbf{A}_0$, is described by the Littlejohn~\cite{littlejohn1983variational} Lagrangian,
\begin{align}
\label{eq:Littlejohn_equilibrium}
\begin{split}
L_0(\mathbf{R},\mathbf{\dot R},v_{||},t)&=\left(q_{\rm i}\mathbf{A}_0+\frac{M_{\rm i} v_{||}}{B_0}\mathbf{B}_0\right)\cdot\mathbf{\dot R}-\frac{M_{\rm i}v_{||}^2}{2}-\mu_i B_0,
\end{split}
\end{align}
where $t$ is time, $\mathbf{R}$ is the position of the particle, $\mathbf{\dot R}$ is the velocity of the particle, $v_{||}=\mathbf{v}\cdot\mathbf{B}_0/B_0$ is the particle velocity along the equilibrium magnetic field, and $\mu_i=M_{\rm i}(v^2-v_{||}^2)/(2B_0)$ is the conserved magnetic moment of the particle.
CATAPULT solves the equations of motion in \Cref{eq:Littlejohn_equilibrium} in both Boozer and Cartesian coordinates
using the Dormand-Prince adaptive timestepping scheme.
Using Cartesian coordinates allows for tracing past the last closed flux surface as well as
the presence of islands in the magnetic field and has been implemented in SIMSOPT \cite{landreman2021simsopt}, ASCOT \cite{hirvijoki2014ascot}, and TAPaS \cite{zarzoso2022transport}.
Boozer coordinates $(s, \theta, \zeta, \vpar)$ provide a convenient representation of the equations of motion
in flux coordinates adapted to the magnetic field geometry and are also implemented in SIMSOPT \cite{landreman2021simsopt} and SIMPLE \cite{albert2020accelerated}.
In this case, $s$ is the flux surface label, $\theta$ is the poloidal angle, $\zeta$ is toroidal angle, and $\vpar$ is the component of the velocity parallel to $B$. 
Other representations of charged particle motion exist and we point the interested reader to
Imbert et al. \cite{imbert2024introduction} for a comparison of these coordinate systems and
broader discussion of \Cref{eq:Littlejohn_equilibrium}.

\section{Magnetic Field Representation}

In each coordinate system and model of the plasma (vacuum, finite $\beta$, etc.),
there are several magnetic field quantities that appear in the equations of motion.
For a vacuum plasma in Boozer coordinates $(s, \theta, \zeta, \vpar)$, these quantities are $\modB,
\frac{\partial \modB}{\partial s}, \frac{\partial \modB}{\partial \theta}, \frac{\partial \modB}{\partial \zeta}$,
and $\iota$.
This is represented on a regular interpolant grid in the spatial coordinates
$s, \theta, \zeta$.
Note that the magnetic field quantities depend on the particle's position,
but not its parallel velocity.

Our grid is similar to the one employed by SIMSOPT \cite{landreman2021simsopt} which uses
fixed cells for the interpolant. 
For a given cell, the magnetic field quantities are represented on a $4\times 4 \times 4$ grid of points
covering the cell and we use tricubic interpolation on the grid to recover a dense
representation of the magnetic field.
As a particle moves from one cell to another, the data used in the interpolation changes.
This enforces the continuity of the field quantities, but not necessarily the differentiability.
An example is shown in \Cref{fig:example_spline}.
When tracing in Cartesian coordinates, we use a regular grid in the cylindrical coordinates
$(R, \phi, z)$.

\begin{figure}
\centering
\caption{An example cubic spline in $s$ with two cells. Data is known on $7$ nodes and continuity is guaranteed on the boundary between cells, but differentiability is not.}
\label{fig:example_spline}
\begin{tikzpicture}[
  x=1.05cm, y=1.25cm,
  font=\sffamily\small,
  note/.style={font=\sffamily\scriptsize,black!65},
  pt/.style={circle,draw=black!60,fill=white,inner sep=0pt,minimum size=4pt},
]

\colorlet{cA}{blue!55!black}
\colorlet{cB}{orange!75!black}

% ---------- the two cubics ----------
% cell 0: Lagrange through (0,.6) (1,1.7) (2,2.2) (3,2.4)
\draw[cA,line width=0.9pt,domain=0:3,samples=61,variable=\x] plot (\x,{
    0.6*(-(\x-1)*(\x-2)*(\x-3)/6)
  + 1.7*( (\x  )*(\x-2)*(\x-3)/2)
  + 2.2*(-(\x  )*(\x-1)*(\x-3)/2)
  + 2.4*( (\x  )*(\x-1)*(\x-2)/6)});

% cell 1: Lagrange through (3,2.4) (4,1.2) (5,.9) (6,1.9)
\draw[cB,line width=0.9pt,domain=3:6,samples=61,variable=\x] plot (\x,{
    2.4*(-(\x-4)*(\x-5)*(\x-6)/6)
  + 1.2*( (\x-3)*(\x-5)*(\x-6)/2)
  + 0.9*(-(\x-3)*(\x-4)*(\x-6)/2)
  + 1.9*( (\x-3)*(\x-4)*(\x-5)/6)});

% ---------- grid points ----------
\foreach \i/\y in {0/0.6, 1/1.7, 2/2.2, 4/1.2, 5/0.9, 6/1.9}{
  \draw[black!18,line width=0.3pt,dotted] (\i,0.18) -- (\i,\y);
  \node[pt] at (\i,\y) {};}
\draw[black!25,line width=0.3pt,dotted] (3,0.18) -- (3,2.4);
\node[pt,fill=black!70,minimum size=5pt] at (3,2.4) {};

% ---------- axis with point indices ----------
\draw[black!45,line width=0.4pt] (-0.25,0.18) -- (6.3,0.18)
  node[note,right,black!55] {$s$};

% ---------- cell underbraces ----------
\draw[decorate,decoration={brace,amplitude=4pt,mirror},cA,line width=0.5pt]
  (0.02,-0.13) -- (2.98,-0.13);
\node[note,text=cA] at (1.5,-0.42) {cell 0};
\draw[decorate,decoration={brace,amplitude=4pt,mirror},cB,line width=0.5pt]
  (3.02,-0.13) -- (5.98,-0.13);
\node[note,text=cB] at (4.5,-0.42) {cell 1};

\end{tikzpicture}
\end{figure}
% \begin{figure}
%     \centering
%     \includegraphics[width=0.6\linewidth]{./figures/example_spline.png}
%     \caption{An example cubic spline in $s$ with two cells. Data is known on $7$ nodes and continuity is guaranteed on the boundary between cells, but differentiability is not.
%     Interpolant colored by cell. }
%     \label{fig:example_spline}
% \end{figure}

\section{Implementation}

To describe the implementation of CATAPULT, we start at the top and work downward.
At the highest level, the inputs are magnetic field data on the interpolant grid and
a set of initial positions and parallel velocities along with the mass, charge, and total
energy of the particles being traced.
We assume all particles have the same energy, mass, and charge.
Timesteps for each particle are computed sequentially until $\min(\tau(X_i) , T)$
is reached, where $\tau$ is the time the particle reaches the last closed flux surface.
In Boozer coordinates, the last closed flux surface corresponds to $s \ge 1$ and in Cartesian coordinates
we detect a sign change in the interpolation of a signed distance function on the interpolant grid.
When the kernel is launched, a block of 32 threads (1 warp on NVIDIA GPUs) is assigned a group of 8 particles to trace.
The relevant data for those particles and memory for future intermediate calculations is allocated in shared memory, where it remains for the duration of the simulation, with the exception of intermediate derivative evaluations which are stored in global memory.
This choice is tuned for A100 GPUs, but on other devices with a different amount of shared memory and registers, a different number of particles per block may be optimal.

CATAPULT uses the Dormand-Prince 5 (DP5) adaptive timestepper \cite{dormand1980family}, matching
the implementation in Boost \cite{BoostOdeint} which is used by SIMSOPT \cite{landreman2021simsopt}.
Each timestep has stages with three parts: computing the next DP5 stage, computing the interpolants, and computing the derivatives from the interpolants.
At the end, we accept/reject the timestep using data from all stages.

Building an intermediate stage, where derivatives are evaluated requires mapping all three coordinates into the interpolant grid by exploiting toroidal and stellarator symmetry. 
With 8 threads per block, this is 24 independent tasks, which we assign to independent threads.
Calculating the subsequent derivative is done independently with one thread per particle, using the interpolant results, which are computed cooperatively.
The performance of CATAPULT is largely derived from the performance of the interpolant.
As an example, we will consider tracing particles using the guiding center time derivatives in Boozer coordinates.
This requires interpolating 5 quantities: $\|B\|, \frac{\partial \|B\|}{\partial s}, \frac{\partial \|B\|}{\partial \theta}, \frac{\partial \|B\|}{\partial \zeta}, \iota$.
This corresponds to $5$ inner products with $64$ elements each.
Interpolant data is stored such that each set of $64$ interpolant values are contiguous (\Cref{fig:data_layout}).
Once the interpolant cell is identified, the warp cooperates on the $40$ independent inner products, computing 4 of them simultaneously.

The layout of the interpolant data is redundant at small cost. 
For $k$ cells in each direction and $m$ interpolant values, this corresponds to an interpolant data
array of size $64 k^3 \times m$ because each point in each cell is stored uniquely,
meaning points on the boundary between cells are stored redundantly.
There are $(3k+1)^3$ unique nodes and these could be stored in an array of size $m(3k+1)^3$ at the cost of reduced cache locality but the memory savings are bounded by a factor of 2.
For usual choices of $k$ such as 15 or 30, both representations use less than 1GB of data which is far below the bound of smaller A100s at 40GB. This size can be accommodated on most modern GPUs.

\begin{figure}
\centering
\caption{Interpolant data layout and accesses}
\label{fig:data_layout}
\begin{tikzpicture}[
  font=\sffamily\small,
  note/.style={font=\sffamily\scriptsize,black!65},
]

\def\W{6.8}      % total width
\def\RH{0.45}    % row height
\def\NC{5}       % number of columns (vacuum)

% --- body
\draw[fill=teal!10,draw=black!55] (0,0) rectangle (\W,\NC*\RH);

% 64 value slots
\foreach \c in {1,...,63}{
  \draw[black!12,line width=0.2pt] (\c*\W/64,0) -- (\c*\W/64,\NC*\RH);
}
% heavier separators every 16 (the four i-blocks)
\foreach \c in {16,32,48}{
  \draw[black!45] (\c*\W/64,0) -- (\c*\W/64,\NC*\RH);
}
% row separators
\foreach \r in {1,...,5}{
  \draw[black!35] (0,\r*\RH) -- (\W,\r*\RH);
}

% --- row labels (top to bottom = column order of quad_info)
\foreach \r/\name in {0/{modB},1/{dmodBds},2/{dmodBdtheta},
                      3/{dmodBdzeta},4/{iota}}{
  \node[note,anchor=east] at (-0.12,{\NC*\RH-0.5*\RH-\r*\RH}) {\texttt{\name}};
}

% --- top brace
\draw[decorate,decoration={brace,amplitude=5pt},black!60]
  (0,\NC*\RH+0.12) -- (\W,\NC*\RH+0.12);
\node[note] at (\W/2,\NC*\RH+0.62) {64 values, contiguous in memory};

% --- bottom: index ticks and i-blocks
\node[note] at (\W/128,-0.22) {0};
\node[note] at ({\W-\W/128},-0.22) {63};
% \foreach \b in {0,...,3}{
%   \draw[decorate,decoration={brace,amplitude=3pt,mirror},black!45]
%     ({\b*\W/4+0.03},-0.42) -- ({(\b+1)*\W/4-0.03},-0.42);
%   \node[note] at ({(\b+0.5)*\W/4},-0.75) {$i=\b$};
% }
% \node[note] at (\W/2,-1.20)
%   {slot $=$ \texttt{16*i + 4*j + k}\quad ($\zeta$ index $k$ fastest)};

% --- caption
\node[note] at (\W/2,-1.65)
  {\texttt{interp\_data[cell\_id,:,:]}\; $(5\times 64)$};

\end{tikzpicture}
\end{figure}

Iterating over particles, and interpolant values within a single particle, the warp first loads the data for interpolating $\|B\|$ in $4$ iterations.
On each iteration, two interpolant values are loaded from global memory.
As each value is loaded, it is multiplied by the appropriate interpolation weight and accumulated locally on a thread.
This accumulation is shown in \Cref{fig:warp_access} for interpolating \texttt{modB} for the first 4 particles.
These reads are fully coalesced on an A100 and represent contiguous $64\,B$ reads from global memory, which can be coalesced on any modern NVIDIA GPU. 
Once these elements are summed, the warp iterates over interpolated values (re-using weight computations) and particles within the warp.

\begin{figure*}
\centering
\caption{The first load instruction for interpolating \texttt{modB} across a warp.
Each row is the 64-value \texttt{modB} slice of a different particle's cell;
the number in a cell is the thread that reads it.}
\label{fig:warp_access}
\makebox[\textwidth][c]{%
\begin{tikzpicture}[
  font=\sffamily\small,
  note/.style={font=\sffamily\scriptsize,black!65},
]

\def\W{15}      % width of one 64-value row
\def\RH{0.42}   % row height
\def\VS{0.74}   % vertical pitch between rows
\def\LPAD{1.3}  % left label overhang; mirrored on the right to centre the bar
\pgfmathsetmacro{\CW}{\W/64}

\foreach \p in {0,...,3}{
  \pgfmathsetmacro{\y}{-\p*\VS}
  % row background
  \fill[teal!8] (0,\y) rectangle (\W,{\y+\RH});
  % 64 slot separators
  \foreach \c in {1,...,63}{
    \draw[black!12,line width=0.15pt] ({\c*\CW},\y) -- ({\c*\CW},{\y+\RH});}
  % heavier separators every 16
  \foreach \c in {16,32,48}{
    \draw[black!45,line width=0.5pt] ({\c*\CW},\y) -- ({\c*\CW},{\y+\RH});}
  % the 8 slots this particle's lanes touch
  \foreach \l in {0,...,7}{
    \pgfmathtruncatemacro{\t}{8*\p+\l}
    \filldraw[fill=blue!22,draw=black!45,line width=0.2pt]
      ({\l*\CW},\y) rectangle ({(\l+1)*\CW},{\y+\RH});
    \node[font=\fontsize{4}{4}\selectfont,black!80] at ({(\l+0.5)*\CW},{\y+0.5*\RH}) {\t};}
  \draw[black!60] (0,\y) rectangle (\W,{\y+\RH});
  \node[note,anchor=east] at (-0.14,{\y+0.5*\RH}) {particle \p};
  \node[note,anchor=west] at ({8*\CW+0.10},{\y+0.5*\RH}) {64\,B};
}

% --- brace over the loaded segment
\draw[decorate,decoration={brace,amplitude=4pt},blue!55!black,line width=0.5pt]
  (0.02,{\RH+0.10}) -- ({8*\CW-0.02},{\RH+0.10});
\node[note,text=blue!55!black,anchor=west] at (0,{\RH+0.55})
  {8 consecutive doubles, 128\,B-aligned};

% --- slot ruler on the top row
\node[note] at ({0.5*\CW},{\RH+0.22}) {0};
\node[note] at ({63.5*\CW},{\RH+0.22}) {63};
\node[note,anchor=east] at (-0.14,{\RH+0.55}) {\texttt{modB}};

% --- notes, centred on the bar
\pgfmathsetmacro{\yb}{-3*\VS}
\node[note] at ({0.5*\W},{\yb-0.34})
  {Thread \texttt{t} reads slot \texttt{t\%8} of particle \texttt{t/8}:
   32 lanes, 32 distinct doubles, four 64\,B segments.};
\node[note] at ({0.5*\W},{\yb-0.68})
  {Rows are separate cells and are \emph{not} adjacent in memory, so the four
   segments are independent;};

% --- phantom pad: mirrors the label overhang so the bar sits at true centre
\path (-\LPAD,0) rectangle ({\W+\LPAD},0);

\end{tikzpicture}}
\end{figure*}

\begin{figure*}
\centering
\caption{Warp access pattern over full interpolation of \texttt{modB}.
Each row is the 64-value \texttt{modB} slice of a different particle's cell;
cells are labeled by the thread that accesses them and colored by their read operation.}
\label{fig:warp_access}
\makebox[\textwidth][c]{%
\begin{tikzpicture}[
  font=\sffamily\small,
  note/.style={font=\sffamily\scriptsize,black!65},
]

\def\W{15}      % width of one 64-value row
\def\RH{0.42}   % row height
\def\VS{0.74}   % vertical pitch between rows
\def\LPAD{1.2}  % left label overhang; mirrored on the right to centre the bar
\pgfmathsetmacro{\CW}{\W/64}

\foreach \p in {0,...,3}{
  \pgfmathsetmacro{\y}{-\p*\VS}
  % --- 64 cells: iteration = hue, load instruction = tint
  \foreach \ii/\cA/\cB in {0/blue!14/blue!32, 1/orange!16/orange!36,
                           2/teal!16/teal!36, 3/violet!14/violet!30}{
    \foreach \h/\cc in {0/\cA, 1/\cB}{
      \foreach \l in {0,...,7}{
        \pgfmathtruncatemacro{\s}{16*\ii + 8*\h + \l}
        \pgfmathtruncatemacro{\t}{8*\p + \l}
        \filldraw[fill=\cc,draw=black!18,line width=0.15pt]
          ({\s*\CW},\y) rectangle ({(\s+1)*\CW},{\y+\RH});
        \node[font=\fontsize{4}{4}\selectfont,black!80] at ({(\s+0.5)*\CW},{\y+0.5*\RH}) {\t};}}}
  % --- iteration boundaries
  \foreach \c in {16,32,48}{
    \draw[black!50,line width=0.5pt] ({\c*\CW},\y) -- ({\c*\CW},{\y+\RH});}
  \draw[black!60] (0,\y) rectangle (\W,{\y+\RH});
  \node[note,anchor=east] at (-0.14,{\y+0.5*\RH}) {particle \p};
}

% --- slot ruler on the top row
\node[note] at ({0.5*\CW},{\RH+0.22}) {0};
\node[note] at ({63.5*\CW},{\RH+0.22}) {63};
\node[note,anchor=east] at (-0.14,{\RH+0.22}) {\texttt{modB}};

% --- underbraces: one per iteration
\pgfmathsetmacro{\yb}{-3*\VS}
\foreach \ii/\cd in {0/blue!55!black, 1/orange!70!black,
                     2/teal!60!black, 3/violet!60!black}{
  \draw[decorate,decoration={brace,amplitude=4pt,mirror},draw=\cd,
        line width=0.5pt]
    ({16*\ii*\CW+0.03},{\yb-0.12}) -- ({(16*\ii+16)*\CW-0.03},{\yb-0.12});
  \node[note,text=\cd] at ({(16*\ii+8)*\CW},{\yb-0.48}) {\texttt{iteration }\ii};}

% --- legend
% \filldraw[fill=blue!14,draw=black!45] ({0.5*\W-2.9},{\yb-1.10})
%   rectangle ({0.5*\W-2.64},{\yb-0.88});
% \node[note,anchor=west] at ({0.5*\W-2.56},{\yb-0.99})
%   {\texttt{data[base+64*zz]}};
% \filldraw[fill=blue!32,draw=black!45] ({0.5*\W+0.30},{\yb-1.10})
%   rectangle ({0.5*\W+0.56},{\yb-0.88});
% \node[note,anchor=west] at ({0.5*\W+0.64},{\yb-0.99})
%   {\texttt{data[base+8+64*zz]}};

% --- notes, centred on the bar
\node[note] at ({0.5*\W},{\yb-1.46})
  {Thread \texttt{t} reads two 64\,B segments per iteration per particle,};
\node[note] at ({0.5*\W},{\yb-1.80})
  {Over the four iterations the warp consumes
   all 64 values of each row.};

% --- phantom pad: mirrors the label overhang so the bar sits at true centre
\path (-\LPAD,0) rectangle ({\W+\LPAD},0);

\end{tikzpicture}}
\end{figure*}

Once a warp has loaded and accumulated the \texttt{modB} values for the current set of particles, each thread has locally accumulated $8$ elements of the final sum.
These are reduced to $4$ final sum elements through a warp-level reduction.
We use the \texttt{\_\_shfl\_down\_sync} operator to accumulate the final sums.
This is visualized for the 8 contiguous threads that cooperate on a particle in \Cref{fig:warp_reduction}.

\begin{figure}
\centering
\caption{Warp reduction of one interpolant component. $\Sigma_{S}$ is
$\sum_{\ell\in S}v_\ell$; dashed cells hold values never read again. Lanes
8--15, 16--23 and 24--31 reduce the other three particles identically. The
shuffle is warp-wide, so lanes $\ell\ge4$ read across the group boundary at
\texttt{offset=4}; those results are discarded before reaching $\ell=0$.}
\label{fig:warp_reduction}
\begin{tikzpicture}[
  font=\sffamily\small,
  note/.style={font=\sffamily\tiny,black!65},
  lane/.style={draw=black!55,fill=teal!12,rounded corners=0.8pt,
               minimum width=0.78cm,minimum height=0.40cm,inner sep=0.5pt,
               font=\sffamily\tiny},
  dead/.style={draw=black!20,dashed,fill=black!3,rounded corners=0.8pt,
               minimum width=0.78cm,minimum height=0.40cm,inner sep=0.5pt},
  shf/.style={->,>=Latex,blue!55!black,line width=0.45pt},
  keep/.style={->,>=Latex,black!30,line width=0.35pt},
]

\def\CW{0.95}   % lane pitch
\def\VS{0.95}   % stage pitch
\def\LX{-0.58}  % right edge of the left labels

% ---------- lane header ----------
\foreach \l in {0,...,7}{\node[note] at ({\l*\CW},0.45) {$\ell=\l$};}

% ---------- stage 0 : input ----------
\foreach \l in {0,...,7}{\node[lane] (a\l) at ({\l*\CW},0) {$v_{\l}$};}
\node[note,anchor=east] at (\LX,0) {\texttt{local\_val}};

% ---------- stage 1 : offset = 4 ----------
\foreach \l/\txt in {0/{$\Sigma_{04}$}, 1/{$\Sigma_{15}$},
                     2/{$\Sigma_{26}$}, 3/{$\Sigma_{37}$}}{
  \node[lane] (b\l) at ({\l*\CW},-\VS) {\txt};}
\foreach \l in {4,...,7}{\node[dead] (b\l) at ({\l*\CW},-\VS) {};}
\node[note,anchor=east] at (\LX,-\VS) {\texttt{offset\,=\,4}};
\foreach \l in {0,...,3}{
  \pgfmathtruncatemacro{\src}{\l+4}
  \draw[shf] (a\src.south) to[out=-90,in=90] (b\l.north);
  \draw[keep] (a\l.south) -- (b\l.north);}

% ---------- stage 2 : offset = 2 ----------
\foreach \l/\txt in {0/{$\Sigma_{0246}$}, 1/{$\Sigma_{1357}$}}{
  \node[lane] (c\l) at ({\l*\CW},{-2*\VS}) {\txt};}
\foreach \l in {2,...,7}{\node[dead] (c\l) at ({\l*\CW},{-2*\VS}) {};}
\node[note,anchor=east] at (\LX,{-2*\VS}) {\texttt{offset\,=\,2}};
\foreach \l in {0,1}{
  \pgfmathtruncatemacro{\src}{\l+2}
  \draw[shf] (b\src.south) to[out=-90,in=90] (c\l.north);
  \draw[keep] (b\l.south) -- (c\l.north);}

% ---------- stage 3 : offset = 1 ----------
\node[lane,fill=blue!20] (d0) at (0,{-3*\VS}) {$\Sigma_{0\text{--}7}$};
\foreach \l in {1,...,7}{\node[dead] (d\l) at ({\l*\CW},{-3*\VS}) {};}
\node[note,anchor=east] at (\LX,{-3*\VS}) {\texttt{offset\,=\,1}};
\draw[shf] (c1.south) to[out=-90,in=90] (d0.north);
\draw[keep] (c0.south) -- (d0.north);

% ---------- output ----------
\pgfmathsetmacro{\yo}{-3*\VS-0.42}
\draw[shf] (d0.south) -- ({0},\yo);
\node[note,text=blue!55!black,anchor=west] at ({0.22},{\yo-0.10})
  {$\ell=0$ writes result};

\end{tikzpicture}
\end{figure}

Finally, the adaptive timestepping portion of the code, which accepts and rejects new steps is also parallelized.
For each state element ($s, \theta, \zeta, \vpar$ in Boozer coordinates), we must estimate the error in our timestepping approximation.
This produces $32$ unique sums, which are parallelized over the $32$ threads.
A warp reduction is used to find the maximum error over the $4$ state elements, which is used to adjust the timestep.
The acceptance and rejection of timesteps among the $8$ particles in a block/warp can cause warp divergence both in the acceptance phase and more broadly as particles take a different number of steps to finish tracing.
To adjust this, we employ work-stealing with a global queue.
As one particle finishes tracing, it is immediately replaced in the warp with a fresh particle.
This leads to better throughput by ensuring that warp cycles are used productively, and sub-linear runtime scaling.

\section{Numerical Details}

\subsection{Extrapolation}

A single Dormand-Prince 5 timestep takes the current position, and iteratively proposed states in the future
where derivatives need to be evaluated.
Sometimes when particles come close to the last closed flux surface, the timestepper requires derivative evaluations at points where  $s > 1$.
This issue also arises in SIMSOPT \cite{landreman2021simsopt}, where the interpolant fails 
silently by returning the last derivative calculation result.
At the time of writing, this issue has not been resolved, but a GitHub issue is open
\href{https://github.com/hiddenSymmetries/simsopt/issues/523}{here}.
CATAPULT solves this issue by using the nearest interpolant cell data as if the particle 
was inside the cell.
This can occasionally lead to particle trajectories stalling as they resolve this transition
outside the last closed flux surface and causes some discrepancies between SIMSOPT and CATAPULT.
We acknowledge that this solution is  imperfect, but Boozer coordinates at these points are undefined,
so even with an exact representation of the B field, this issue would arise.
In Cartesian coordinates, this is not relevant because the magnetic field is well defined everywhere,
although the same extrapolation is used if derivatives are needed outside 
of the interpolant domain.

\subsection{Pseudo-Cartesian coordinates}

The adaptive step size of DP5 is tuned with error estimates of the form

\begin{align} \label{eq:error_est}
\textbf{error} = \max_i \frac{|e_i|}{\textbf{atol}_i c_i + \textbf{rtol}_i(|x_i| + \textbf{dt} |\dot x_i|)}
\end{align}
where $e_i$ is the estimated error in component $i$ of the state and $c_i$ is the characteristic scale.
This calculation encounters two issues:
the coordinate singularity at the axis in Boozer coordinates, and the relative scaling of $\vpar$.

In Boozer coordinates, for a fixed $\zeta$ and $s = 0$, all $\theta$ values correspond to the same
point, leading to small errors in physical space appearing as large errors in timestepping.
For example, if the two estimates of the particle position were $(\varepsilon, 0, 0, \vpar)$ and
$(\varepsilon, \pi, 0, \vpar)$, then the error estimate in the $\theta$ component would be $\pi$,
despite these positions being arbitrarily close in physical space for sufficiently small $\varepsilon$.
We instead calculate the error in Boozer coordinates by transforming to a pseudo-Cartesian coordinate system
with the mapping $(s, \theta, \zeta, \vpar) \to (s\cos\theta, s\sin\theta, \zeta, \vpar)$.
In this case, the worst error estimates would become $2\varepsilon$ which shrinks with $\varepsilon$.
We use this system for tracing particles in Boozer coordinates and estimating errors for single timesteps.

The relative scaling of $\vpar$ appears in both Cartesian and Boozer tracing because the scale of $\vpar$
is much larger than the scale of the other coordinates in MKS units. 
Boost, which is used in SIMSOPT, implements the DP5 timestepper where the same $\textbf{atol}$ value
must be used for all elements.

In CATAPULT, we multiply $\textbf{atol}$ by $v_{\text{total}}$ to account for this scaling and not restrict
the tracing in this way.
Equivalently, we could rescale $\vpar$ by $v_{\text{total}}$ in the equations of motion, which
is implemented in the CPU tracing in \textbf{firm3d}, available on GitHub \href{https://github.com/ColumbiaStellaratorTheory/firm3d}{here} \cite{firm3d2025}. 

\section{Results}

On an A100 SXM4 GPU with 40GB of memory on Perlmutter, our particle tracing kernel uses 88 registers per thread, to obtain 39.43\% of Speed of Light Compute Throughput and 70.34\% Speed of Light Memory Throughput.
Our kernel achieves 29.61\% occupancy with a theoretical occupancy of 31.25\%.
We present runtime results and scaling on several example magnetic fields from Paul et al. \cite{paul2022energetic}.
In particular, we run on the ATEN, HSX, NCSX, QA, and QH devices scaled to ARIES-CS minor radius and field
strength.
We show scaling as the simulation fidelity increases through increased grid resolution,
decreased tracing tolerances (\textbf{atol} and \textbf{rtol} in \Cref{eq:error_est}),
and increasing the number of particles.
All simulations trace particles until they reach the last closed flux surface or $T = 10^{-3}$ seconds.

\begin{figure}  \centering
\includegraphics[scale=0.5]{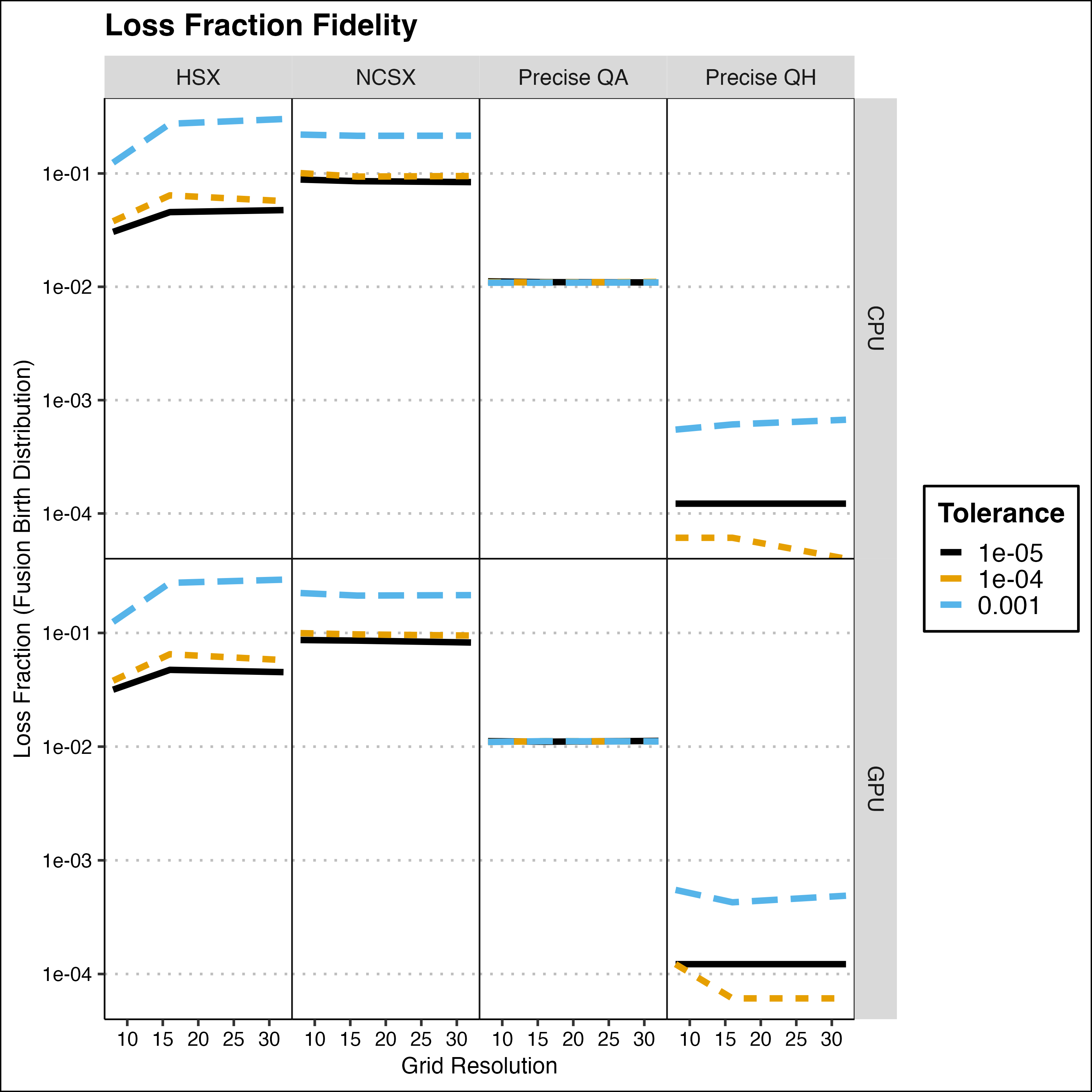}
  \caption{Resolution of loss fraction with varying fidelity for 32,768 particles using guiding center vacuum tracing in Boozer coordinates. Runs with tolerances lower than $10^{-5}$ have been removed because they are visually indistinguishable from the lines corresponding to tolerances of $10^{-5}$.}
\label{fig:unperturbed_loss_frac}
\end{figure}

We first show the consistency between CPU and GPU tracing in \Cref{fig:unperturbed_loss_frac}.
There is a line on each subplot corresponding to each tolerance in the legend, but due to the convergence in fidelity, we observe overlapping lines.
It is also of note that lower fidelity runs (higher tracing tolerances) correspond to higher loss fraction estimates.
Once the grid resolution (number of interpolant cells in each coordinate) reaches 20 and tracing tolerance is $10^{-4}$ or below we see good convergence in the estimated loss fraction with 32,768 particles.

\begin{figure}
  \centering
    \includegraphics[scale=0.45]{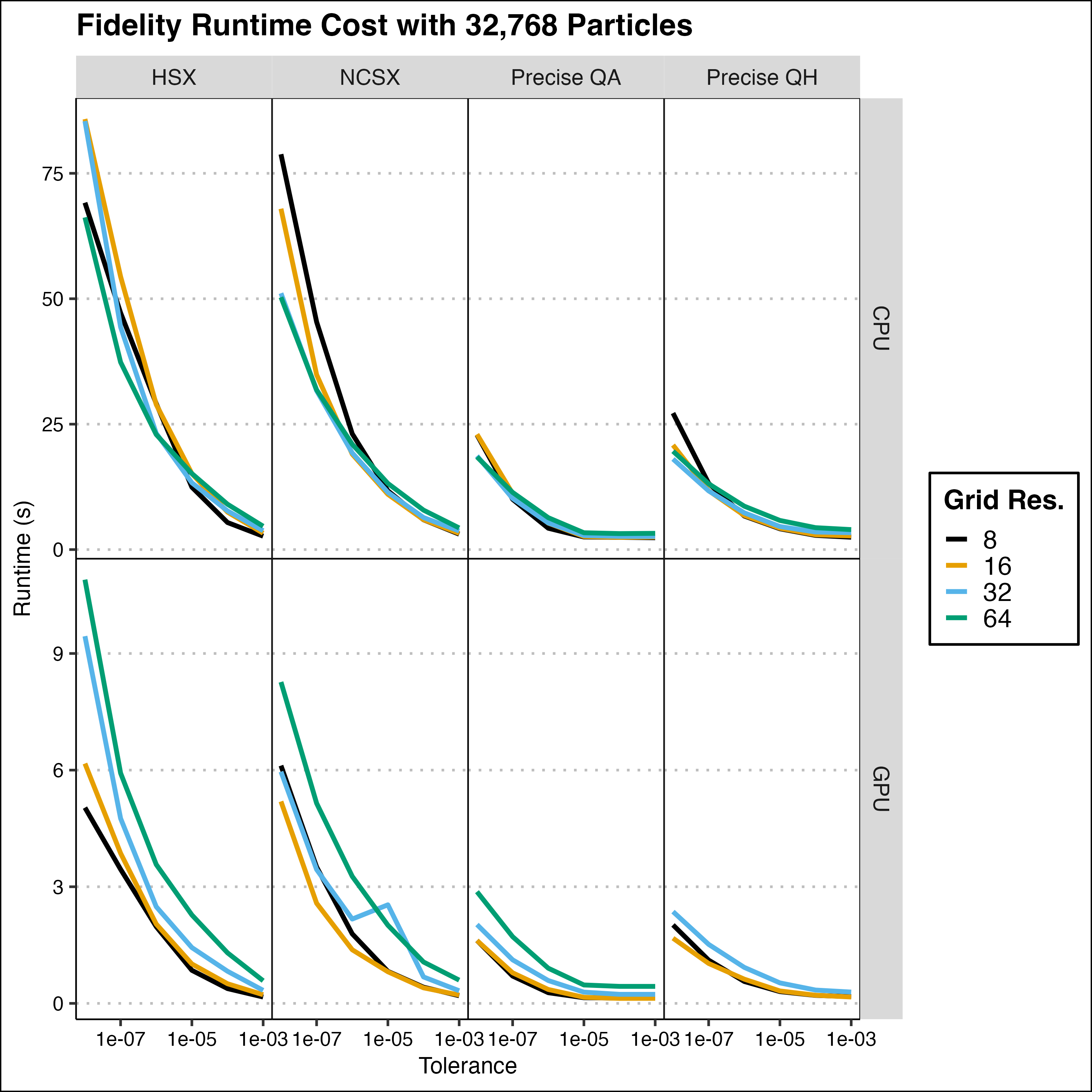}
  \caption{Runtime with varying fidelity. We observe linear scaling  in the number of particles in all cases with longer runtimes when using lower timestep tolerances. Longer runtimes associated with higher fidelity is predominantly caused by lower tracing tolerances as opposed to interpolant grid resolution.}
\label{fig:unperturbed_wallclock}
\end{figure}

\begin{figure}
  \centering
\includegraphics[scale=0.4]{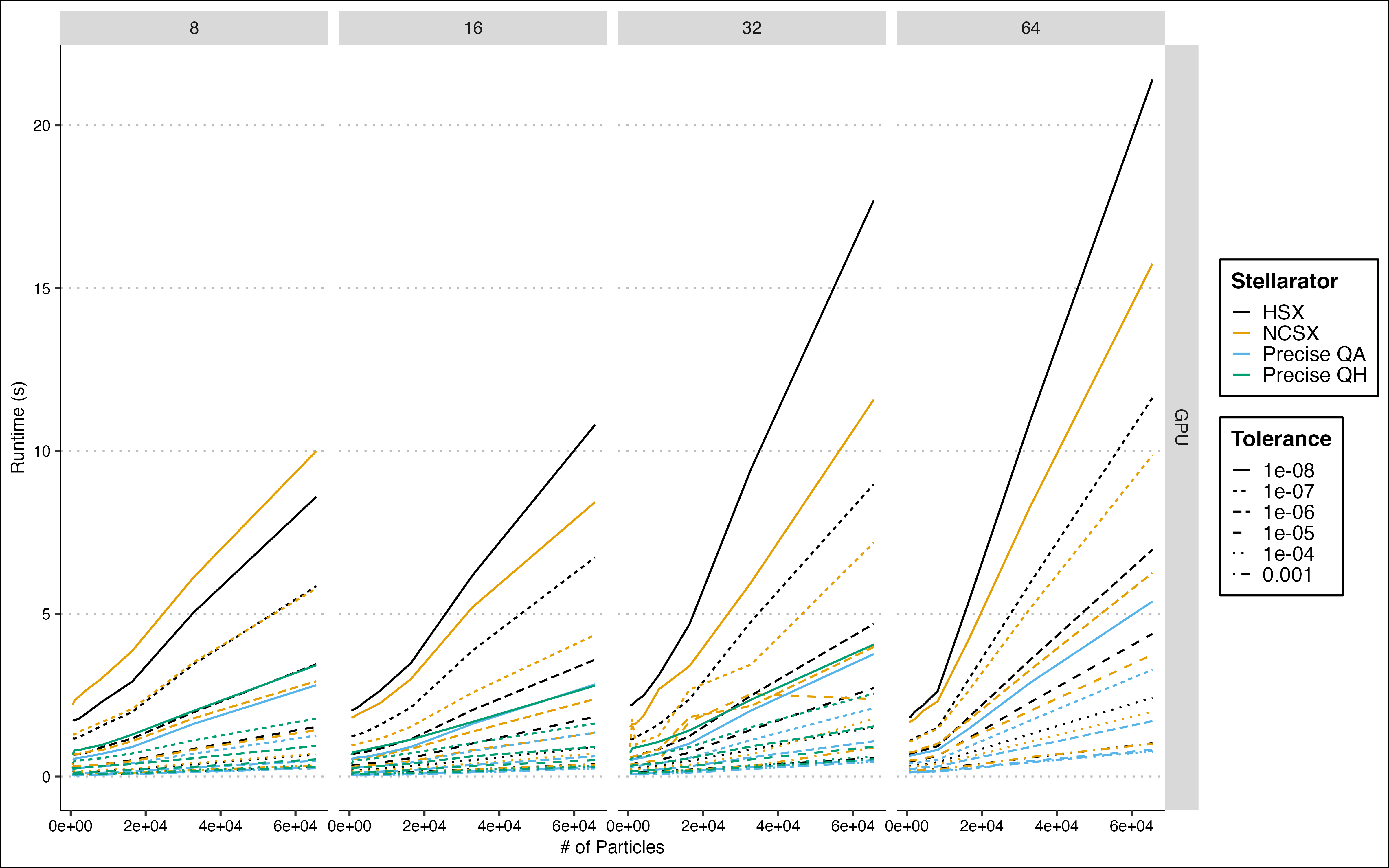}
  \caption{GPU strong scaling for unperturbed vacuum Boozer tracing with varying interpolant grid resolution.}
  \label{fig:unperturbed_strong_scaling}
\end{figure}

\begin{figure}
  \centering
\includegraphics[scale=0.4]{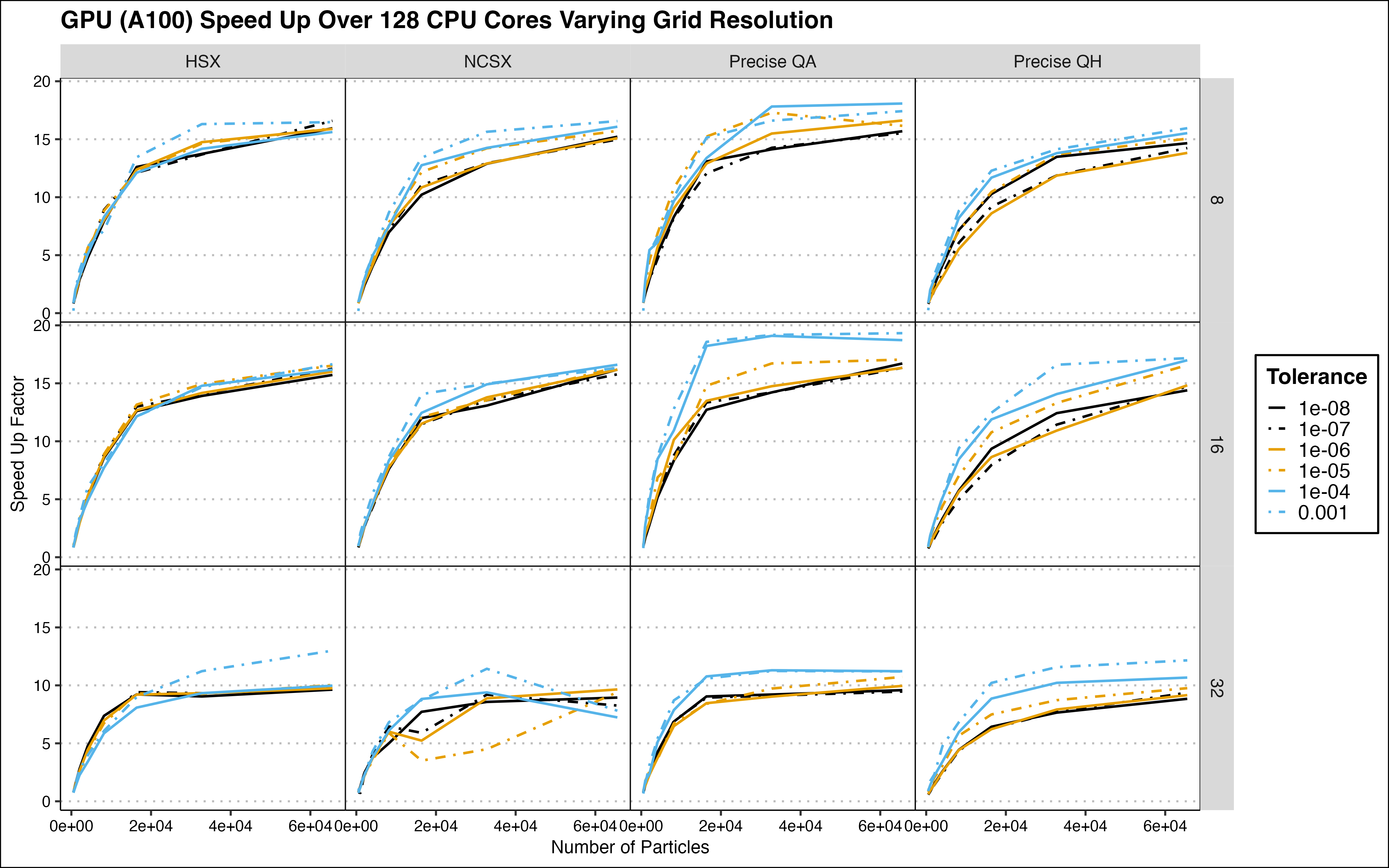}
  \caption{GPU speed up vs. 128 CPU cores on Perlmutter. For all devices, tolerances, and grid resolutions the A100 is saturated around $25,000$ particles at which point the speedup reaches at least a factor of $5$.}
  \label{fig:unperturbed_scaling}
\end{figure}

The runtime cost associated with higher fidelity tracing is presented in \Cref{fig:unperturbed_wallclock}.
Runtime is dominated by increased fidelity through tracing tolerance, and does not seem to be impacted much by grid resolution due to the cache locality of the interpolation scheme for both CPU and GPU tracing.
Note that the scaling of the y axis is different for the CPU and GPU plots.
For a fixed tracing tolerance and grid resolution, we show the speedup of CATAPULT vs. 128 CPU threads on Perlmutter in \Cref{fig:unperturbed_scaling} and raw runtime in \cref{fig:unperturbed_strong_scaling}, both in double precision.
The comparison on Perlmutter is appropriate because many stellarator design tasks are performed on that system. We compare one GPU to 128 CPU threads because this is one CPU node, even though a Perlmutter GPU node contains 4 GPUs. Further scaling across nodes or across GPUs would be linear in time, and consistent with the results we present here.
As the GPU becomes saturated with an increasing number of particles, we observe speedups between a factor of 5 and 10 in each device across grid resolutions and tracing tolerances.

\begin{figure}
  \centering
    \includegraphics[scale=0.4]{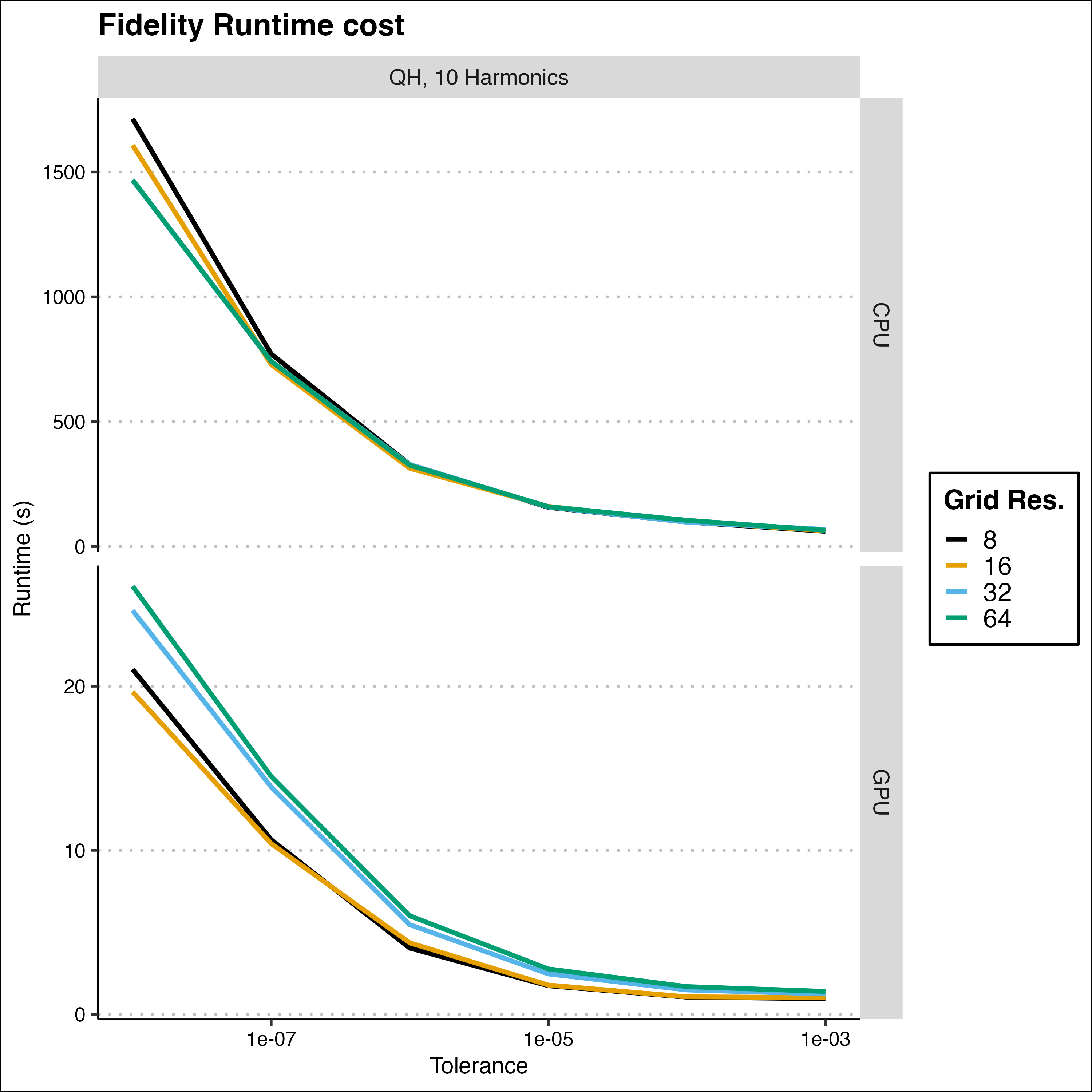}
  \caption{Runtime with varying fidelity for SAWs. We observe linear scaling behavior in the number of particles and only present results with $32,768$ particles. We again observe runtime being impacted by timestep tolerance more than interpolant grid resolution.}
\label{fig:perturbed_wallclock}
\end{figure}

\begin{figure}
  \centering
\includegraphics[scale=0.4]{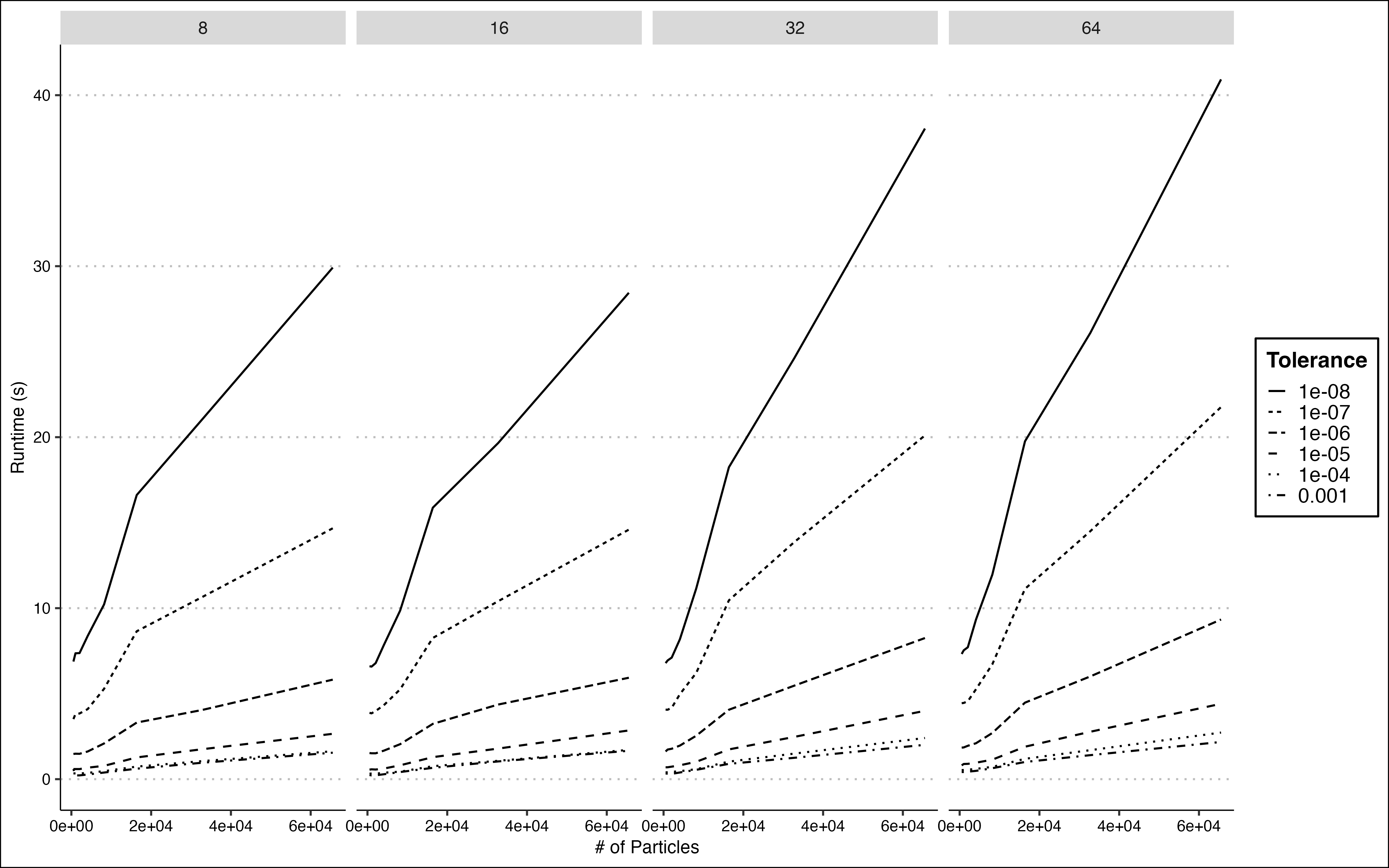}
  \caption{GPU strong scaling for tracing in the presence of Shear Alfven Waves with varying interpolant grid resolution.}
  \label{fig:perturbed_strong_scaling}
\end{figure}

\begin{figure}
  \centering
\includegraphics[scale=0.4]{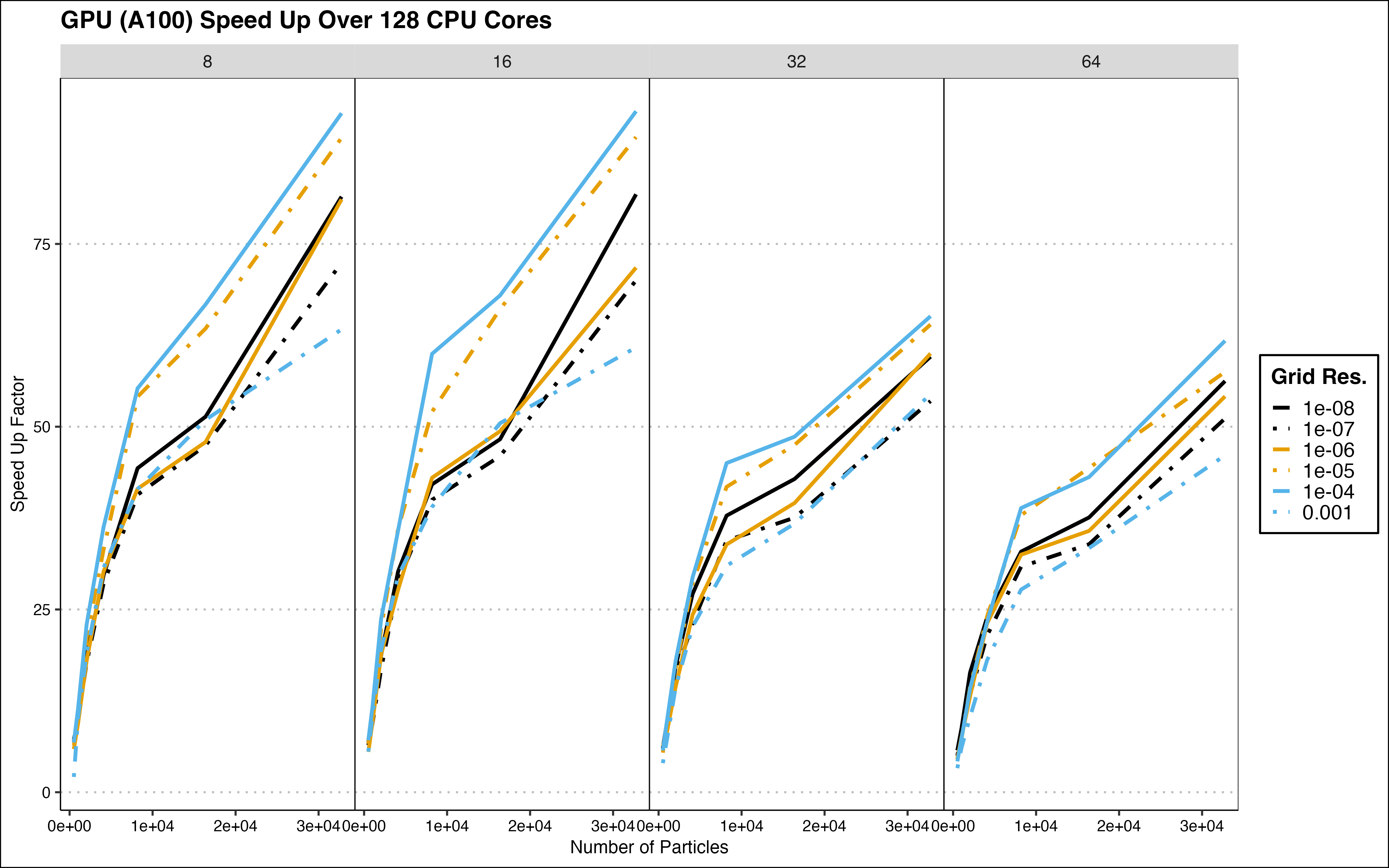}
  \caption{GPU speed up vs. 128 CPU cores on Perlmutter for SAWs. We observer a much larger speed up compared to the unperturbed case.}

  \label{fig:perturbed_scaling}
\end{figure}

Similar plots for vacuum tracing in the presence of Shear Alfven Waves are shown in \Cref{fig:perturbed_wallclock} and \Cref{fig:perturbed_scaling} using the Landreman-Buller QH configuration \cite{landreman2022optimization} using the ideal Alfven equation as in Knyazev et al. \cite{knyazev2026shear}. Raw runtime is shown in \cref{fig:perturbed_strong_scaling}.
The same conclusions hold for the scaling with respect to fidelity, and we observe speed ups on the order of 40 to 60 times faster than the CPU node on Perlmutter.

\begin{figure}
  \centering
    \includegraphics[scale=0.4]{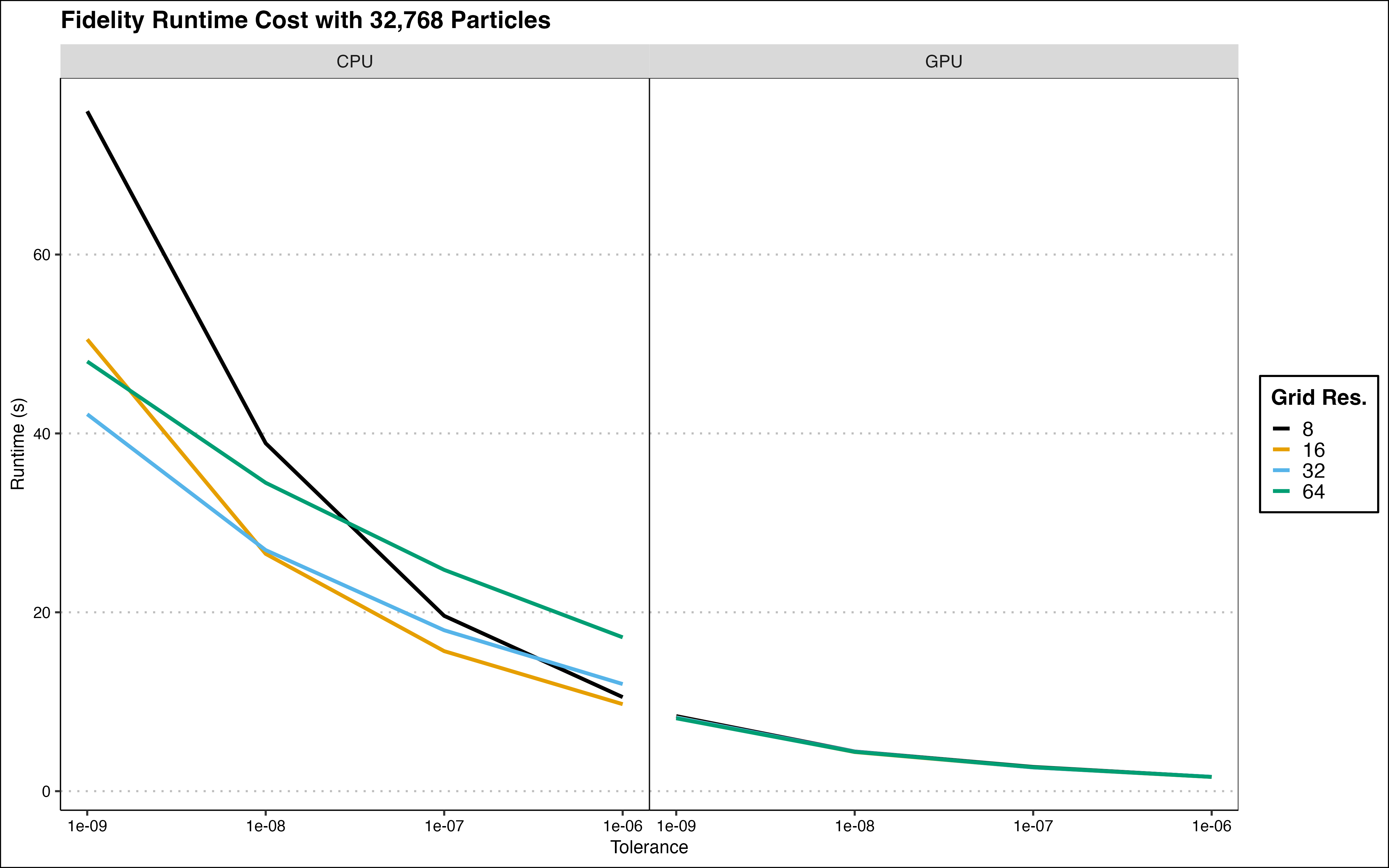}
  \caption{Runtime with varying fidelity for Cartesian vacuum tracing. We observe similar results to vacuum Boozer tracing. Note that on the GPU the lines are overlapping because we observe no runtime cost to increasing the interpolant grid density.}
\label{fig:cartesian_wallclock}
\end{figure}

\begin{figure}
  \centering
\includegraphics[scale=0.3]{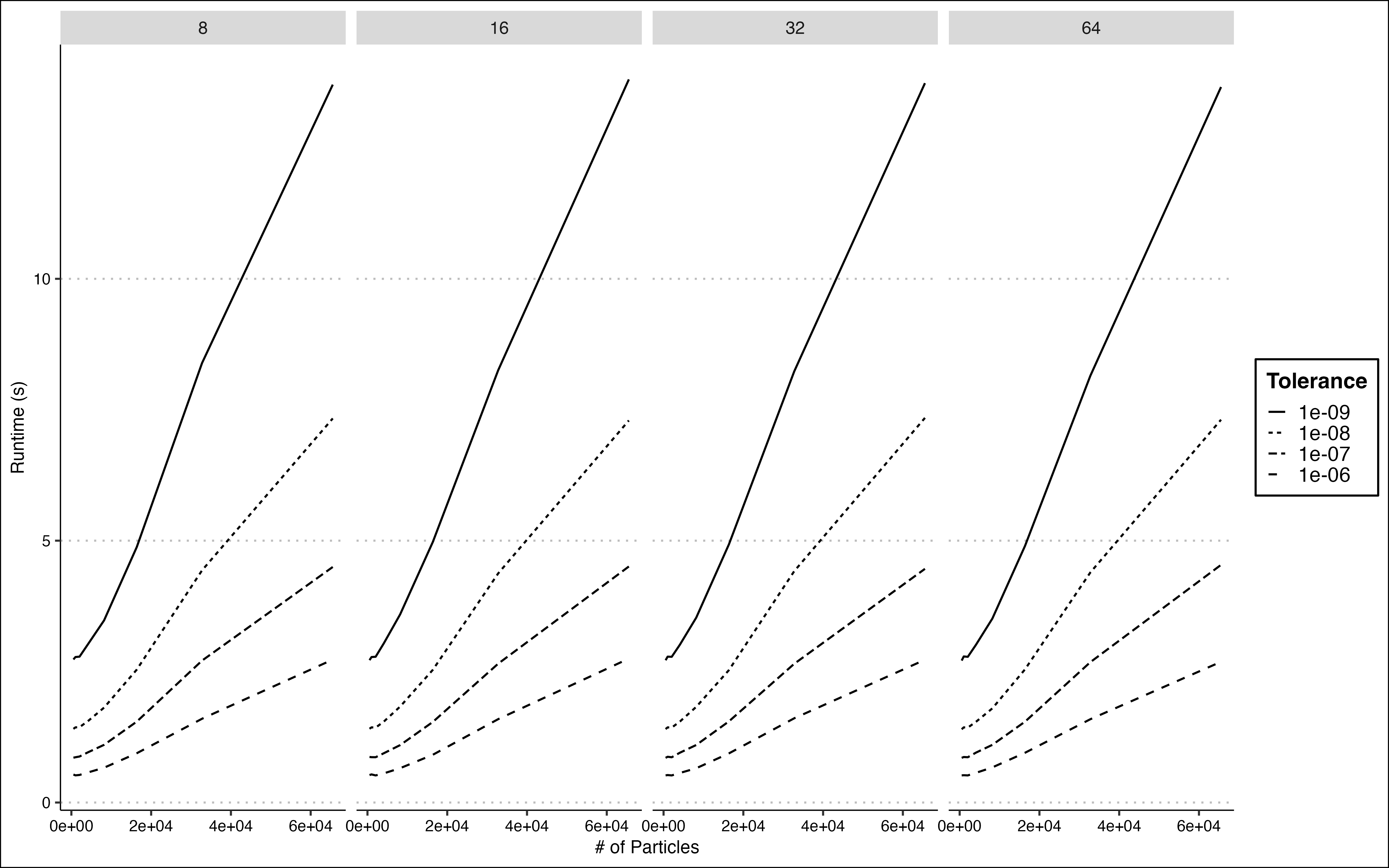}
  \caption{GPU strong scaling with varying interpolant grid resolution.}
  \label{fig:cartesian_strong_scaling}
\end{figure}

\begin{figure}
  \centering
\includegraphics[scale=0.3]{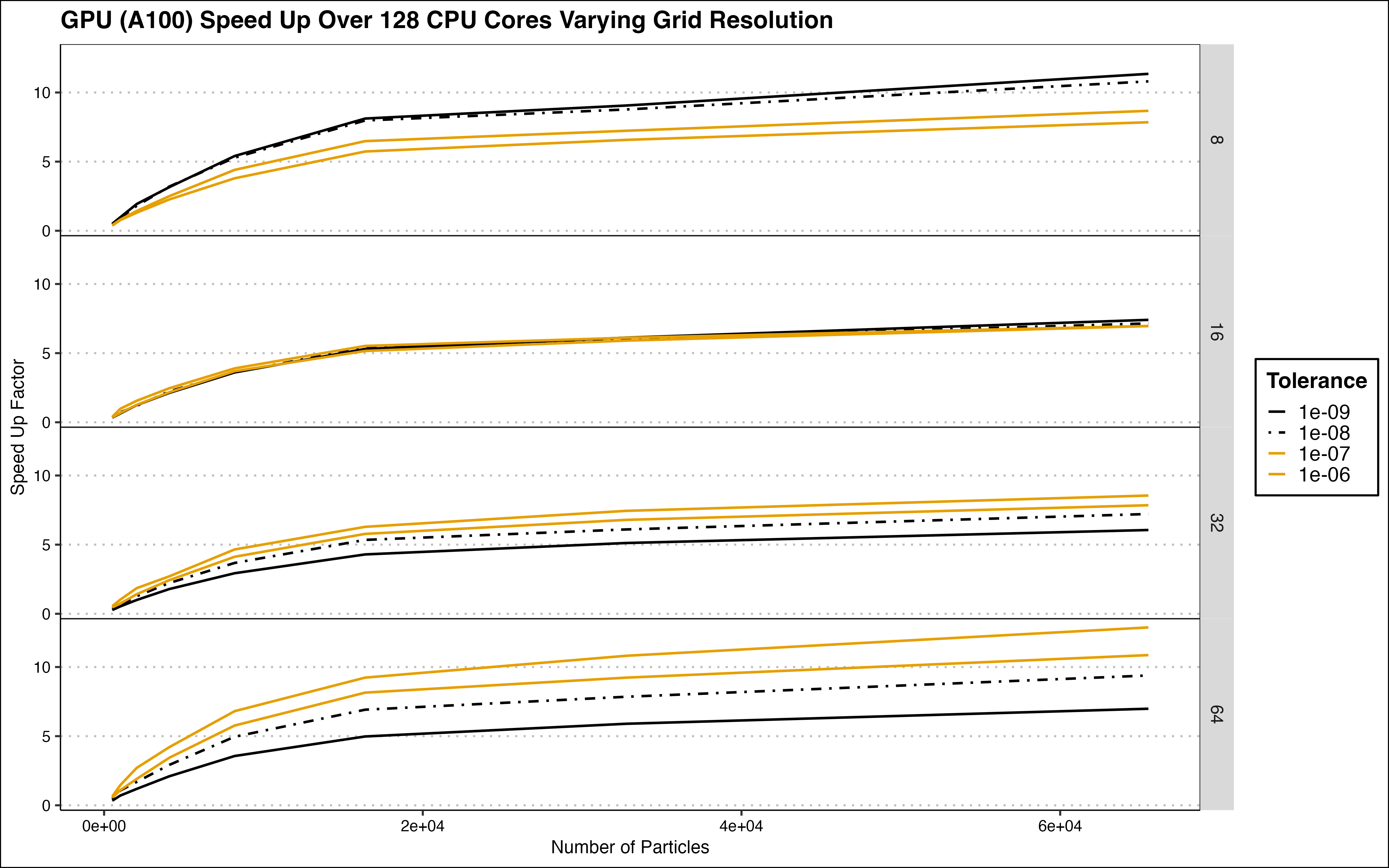}
  \caption{GPU speed up vs. 128 CPU cores on Perlmutter for Cartesian vacuum tracing. The speed up is comparable to the vacuum Boozer case.
  Grid resolution is measured by $n$ where the interpolant grid in $r, \phi, z$ is of size $n, 2n, \frac{n}{2}$ respectively.}
  \label{fig:cartesian_scaling}
\end{figure}

We also present similar results for vacuum Cartesian tracing of the Precise QH configuration in \Cref{fig:cartesian_wallclock,fig:cartesian_scaling,fig:cartesian_strong_scaling}. Particles are launched from the $s=0.3$ surface and traced until they leave the last closed flux surface or $T=10^{-3}$. 
We do not present the loss fractions visually because for all settings they are identically $0$.

Across interpolant grid resolutions, timestep tolerances, and coordinate systems, CATAPULT is on average at least $5$ times faster than 128 CPU threads on a single Perlmutter node. 
We also note that CATAPULT threads share a single copy of the interpolant grid data, where parallelized CPU implementations of SIMSOPT and firm3d copy the interpolant grid to each thread.
By sharing a single copy, CATAPULT is able to scale to higher grid resolutions without running out of memory, enabling much higher fidelity particle tracing.

\section{Conclusion and Future Work}

We have introduced a novel CUDA accelerated GPU tracing implementation for alpha particles in stellarators.
The speedup over existing CPU implementations is consistently $>5$X in various vacuum configurations in both Cartesian and Boozer coordinates.
This enables higher fidelity tracing without a large runtime cost by increasing grid resolution.

The higher throughput of particle tracing allows us to consider more complicated physics, such as full orbit tracing or collisions, and opens the door to new work that requires more samples, such as studying wall loads for alpha particles or directly optimizing confinement.

There is also the possibility of
tuning to other GPUs other than A100s, porting to other GPU libraries, or optimizing for new generations for hardware.
We could also investigate improved interpolation and timestepping schemes to improve the numerical accuracy of our results as well as the speed of our code.
The implementation of CATAPULT leaves ample avenues for further exploration of charged particle motion and understanding of stellarator design.

\section{Acknowledgements}

We acknowledge funding through the U.S. Department of Energy, under contracts
DE-SC0024630, DE-SC0024548, DE-AC02-09CH11466, DE-AC52-07NA27344. We also acknowledge funding
through the Simons Foundation collaboration “Hidden Symmetries and Fusion Energy,”
Grant No. 601958. This research used resources of the National Energy Research Scientific
Computing Center (NERSC), a Department of Energy Office of Science User Facility
using NERSC award ERCAP0031926.

\bibliographystyle{elsarticle-num}
\bibliography{main}

\end{document}